\documentclass{aa}

\usepackage[english]{babel}
\usepackage{epsfig}
\usepackage{longtable}
\usepackage{deluxetable}
\usepackage{lscape}

\newcommand{\CL}{\object{RX~J0152.7-1357}}

\newbox\grsign      \setbox\grsign=\hbox{$>$}
\newdimen\grdimen   \grdimen=\ht\grsign
\newbox\simgreatbox \setbox\simgreatbox=\hbox{\raise.5ex\hbox{$>$}\llap
                        {\lower.5ex\hbox{$\sim$}}}\ht1=\grdimen\dp1=0pt
\newbox\simlessbox  \setbox\simlessbox =\hbox{\raise.5ex\hbox{$<$}\llap
                        {\lower.5ex\hbox{$\sim$}}}\ht2=\grdimen\dp2=0pt
\def\simgreat{\mathrel{\copy\simgreatbox}}
\def\simless {\mathrel{\copy\simlessbox }}
 
\begin{document}

\title{A VLT spectroscopic survey of RX~J0152.7-1357, a forming
cluster of galaxies at z=0.837. \thanks{Based in part on observations
carried out at the European Southern Observatory using the ESO Very
Large Telescope on Cerro Paranal (ESO programs 166.A-0701, 69.A-0683
and 72.A-0759) and the ESO New Technology Telescope on Cerro La Silla
(ESO progrm 61.A-0676).}}

\author{
  R.~Demarco \inst{1,2}
\and
  P.~Rosati \inst{1}
\and
  C.~Lidman \inst{3}
\and
  N. L. ~Homeier \inst{2}
\and
  E.~Scannapieco \inst{4}
\and
  N.~Ben\'{i}tez \inst{2}
\and
  V.~Mainieri \inst{5}
\and
  M.~Nonino \inst{6}
\and
  M.~Girardi \inst{7}
\and
  S. A.~Stanford \inst{8,9}
\and
  P.~Tozzi \inst{6}
\and
  S.~Borgani \inst{10,11}
\and
  J.~Silk \inst{12}
\and
  G.~Squires \inst{13}
\and
  T.~J.~Broadhurst \inst{14}
}
\offprints{R. Demarco (\sl{demarco@pha.jhu.edu}) }
\institute{
        ESO-European Southern Observatory, Karl-Schwarzschild Str. 2,
 D-85748 Garching b. M\"unchen, Germany
\and
        Department of Physics and Astronomy, Johns Hopkins University, 3400 N. Charles Str.,
 Baltimore, MD 21218, USA  
\and
        ESO-European Southern Observatory, Alonso de Cordova 3107, Casilla 19001, Santiago, Chile
\and
        Kavli Institute for Theoretical Physics, Kohn Hall, University of California, Santa Barbara, CA 93106, USA
\and
        Max-Planck-Institut f\"ur extraterrestrische Physik,  Giessenbachstrasse,  
 D-85748 Garching b. M\"unchen, Germany
\and
        INAF - Osservatorio Astronomico di Trieste, via G. B. Tiepolo 11, 34131 Trieste, Italy
\and
        Dipartimento di Astronomia, Universit\`a degli Studi di Trieste, via Tiepolo 11, 34131 Trieste, Italy
\and
        Department of Physics, University of California, Davis, 1 Shields Avenue, Davis, CA 95616, USA
\and
        Participating Guest, Institute of Geophysics and Planetary Physics, Lawrence Livermore National Laboratory, USA
\and
        Dipartimento di Astronomia dell'Università di Trieste, via Tiepolo 11, I-34131 Trieste, Italy
\and
        INFN - National Institute for Nuclear Physics, Trieste Università, via Valerio 2, I-34127 Trieste, Italy
\and
        Astrophysics Department, University of Oxford, Keble Road, Oxford OX1 3RH
\and
        SIRTF Science Center, California Institute of Technology, Pasadena, CA 91125, USA 
\and
        Racah Institute of Physics, Hebrew University, Jerusalem 91904, Israel.
}

\date{}

\authorrunning{R. Demarco et al.}

\titlerunning{A VLT spectroscopic survey of \CL}

\abstract{We present the results of an extensive spectroscopic survey
of \CL, one of the most massive distant clusters of galaxies
known. Multi-object spectroscopy, carried out with FORS1 and FORS2 on
the ESO Very Large Telescope (VLT), has allowed us to measure more
than 200 redshifts in the cluster field and to confirm 102 galaxies as
cluster members. The mean redshift of the cluster is $z=0.837 \pm
0.001$ and we estimate the velocity dispersion of the overall
cluster galaxy distribution to be $\sim 1600 \ \mathrm{km \ s^{-1}}$.
The distribution of cluster members is clearly irregular, with two
main clumps that follow the X-ray cluster emission mapped by
Chandra. A third clump of galaxies to the east of the central
structure and at the cluster redshift has also been identified. The
two main clumps have velocity dispersions of $\sim919$ and $\sim737 \
\mathrm{km \ s^{-1}}$ respectively, and the peculiar velocity of the
two clumps suggests that they will merge into a single more massive
cluster. A segregation in the star formation activity of the member
galaxies is observed. All star forming galaxies are located outside
the high-density peaks, which are populated only by passive
galaxies. A population of red galaxies (belonging to the cluster red
sequence) with clear post-starburst spectral features and [OII]
($\lambda$3727) emission lines is observed in the outskirts of the
cluster. Two AGNs, which were previously confused with the diffuse
X-ray emission from the intracluster medium in ROSAT and BeppoSAX
observations, are found to be cluster members.

 }

\maketitle

\section{Introduction}

\CL\ was discovered in the ROSAT Deep Cluster Survey (RDCS; Rosati et
al. 1998, Della Ceca et al. 2000) as an extended double core X-ray
source in the ROSAT PSPC field rp600005n00 observed in January
1992. It was independently discovered in the WARPS Survey (Ebeling et
al., 2000) and reported in the Bright SHARC survey (Romer et
al. 2000). Spectroscopy of a small number of member galaxies, carried
out with EFOSC1 at the ESO 3.60-m telescope at the Cerro La Silla
Observatory in November 1996, spectroscopically confirmed the cluster
and gave a cluster redshift of $z=0.83$ (Della Ceca et al., 2000),
placing \CL\ among the most distant and X-ray luminous clusters of
galaxies known, akin to \object{MS1054-03} (Donahue et al., 1998; van
Dokkum et al., 1999) and \object{RX J1716.6+6708} (Gioia et al. 1999)
at $z=0.83$ and z=$0.81$ respectively.

The ROSAT observations were used to derive a cluster X-ray
luminosity\footnote{Throughout this paper we assume a $\Lambda$CDM
cosmology with $H_0=70 \ \mathrm{km \ s^{-1} \ Mpc^{-1}}$,
$\Omega_M$=0.3 and $\Omega_{\Lambda}$=0.7.} of $L_X = (7.4 \pm 0.7)
\times 10^{44}\ \mathrm{erg \ s^{-1}}$ in the 0.5-2 keV band, and
BeppoSAX observations yielded a gas temperature of
$kT=6.46^{+1.74}_{-1.19}$ keV and a metallicity of
$0.53^{+0.29}_{-0.24}$ $Z_{\odot}$ (Della Ceca et al. 2000). Recent
observations with Chandra also show the double core structure in the
intra-cluster medium (ICM) and provide some evidence of a possible
merger in progress (Maughan et al. 2003; Huo et al. 2004). These two
main structures, one to the north-east and the other to the
south-west, are separated by $\sim$ 1\farcm6 (corresponding to 730 kpc
at the cluster redshift) on the plane of the sky. The spectroscopic
analysis of the Chandra X-ray data combined with the latest Chandra
calibrations give temperatures of $6.7^{+1.2}_{-1.0}$ and
$8.7^{+2.4}_{-1.8}$ keV and metallicities of $0.17^{+0.19}_{-0.16}$
and $<$0.22 Z$_{\odot}$ for the northern and southern clumps,
respectively (Balestra et al., in preparation).  These temperature
values are consistent, within the error bars, with the BeppoSAX
estimates and those reported in Ettori et al. (2004); however, the
BeppoSAX metallicity (Della Ceca et al., 2000) seems to be somewhat
higher than the new Chandra measurements. The X-ray temperature
measurements of \CL\ are also consistent with Sunyaev-Zeldovich effect
observations of the cluster (Joy et al. 2001).

Most of the {earlier} work on \CL\ has centered on characterizing its
X-ray properties and the thermodynamical state of the hot
intra-cluster gas. In this paper, we present the results of an
extensive spectroscopic survey aimed at identifying a large number of
galaxies belonging to the cluster. We compare our results with the
X-ray information that is publicly available and with a recent weak
lensing analysis of the cluster based on HST/ACS data (Jee et
al. 2004), and we discuss the dynamical state of the
cluster. Ground-based multi-band photometry is also used to study the
distribution of cluster galaxies in colour-magnitude and colour-colour
space. In forthcoming papers, we will combine the spectroscopic
information presented here with HST/ACS data in a detailed study of
the galaxies in \CL.

\section{Spectroscopic Survey}

\subsection{Imaging observations}

\begin{center}
\begin{figure*}[t]
\centerline{\psfig{file=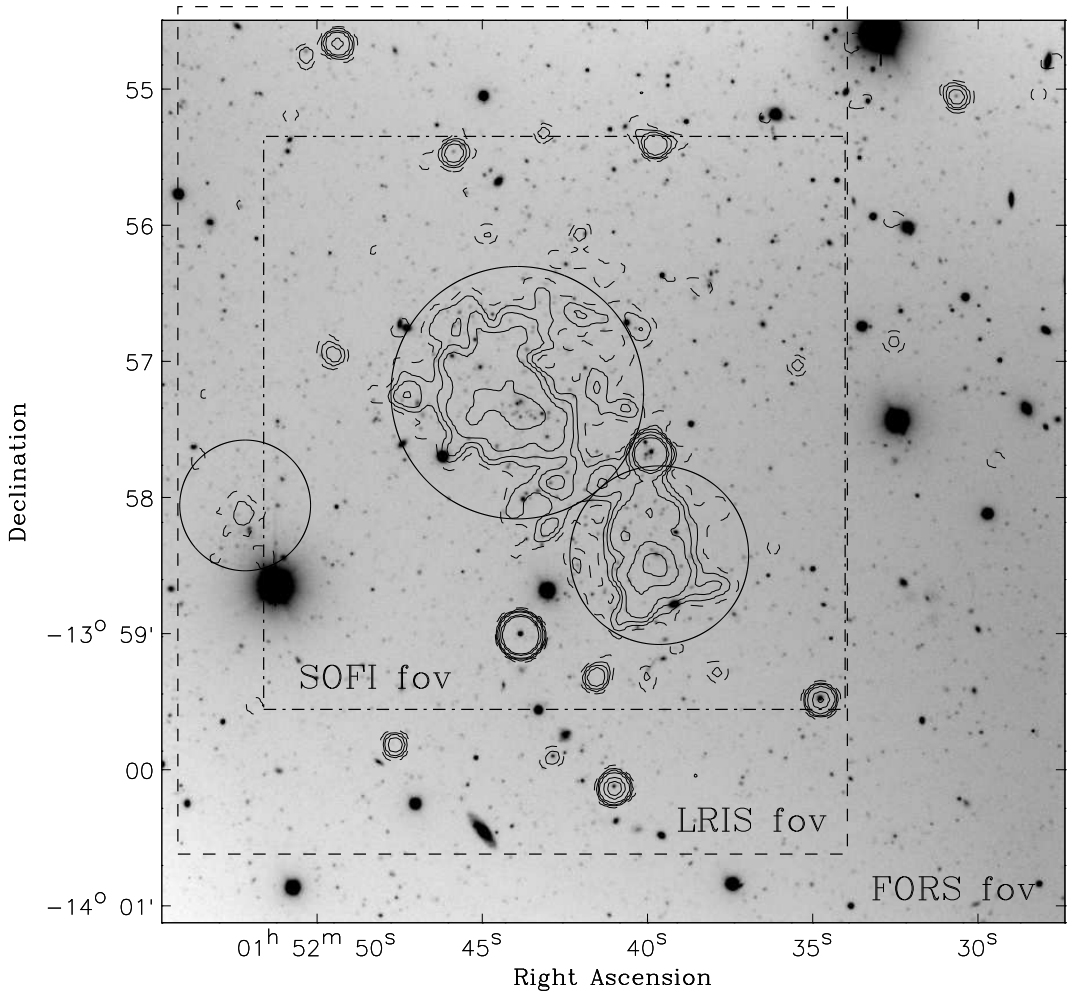,width=15cm}}
\caption{A FORS1 V-, R-, and I-band composite image centered on
\CL. The dashed rectangle corresponds to the 4\farcm9 $\times$
6\farcm54 region covered by the LRIS data superimposed onto the
6\farcm8 $\times$ 6\farcm8 field of view (FOV) of FORS1. The
dot-dashed square corresponds to 4\farcs9 $\times$ 4\farcs9 region
covered by SofI. North is up and East is to the left. Target galaxies
selected from the photometric catalog are within the area delimited by
the dashed rectangle. Also shown are X-ray Chandra iso-contours,
tracing the intracluster medium (ICM) of \CL, as defined in
Fig. \ref{gal_x_dist}. The dashed contour corresponds to the 3$\sigma$
level above the background. Three circular regions, corresponding to
regions of extended X-ray emission, have been defined. The radius of
each region was defined so that most of the X-ray light within the
3$\sigma$ iso-contours is contained in the corresponding region (see
text for details).}
\label{lrisonfors}
\end{figure*}
\end{center}

We used multi-band optical and near-IR imaging observations of \CL\ to
select targets for spectroscopy.  Optical images in the B-, V-, R- and
I-bands were obtained with the Low Resolution Imaging Spectrometer
(LRIS; Oke et al. 1995) at the W. M. Keck Observatory. The LRIS images
cover a region of 4\farcm9 $\times$ 6\farcm54 (see
Fig. \ref{lrisonfors}) with a pixel scale of 0\farcs21 pix$^{-1}$. The
seeing, as measured by the FWHM of points sources, was 0\farcs86 in V,
1\farcs19 in B, 1\farcs0 in R and 0\farcs73 in I. Near-IR images in
the J- and $\mathrm{K_s}$-bands were obtained with SofI (Moorwood,
Cuby \& Lidman 1998) on the ESO NTT telescope at the Cerro La Silla
Observatory. The seeing was 0\farcs95 in J and 0\farcs94 in
$\mathrm{K_s}$. The SofI images cover a region of 4\farcm9 $\times$
4\farcm9 (see Fig. \ref{lrisonfors}) with a pixel scale of 0\farcs29
pix$^{-1}$. The LRIS and SofI images provided the basis for the
photometric catalog used to select target for spectroscopy. The
catalog was created with SExtractor (Bertin \& Arnouts, 1996),
yielding aperture photometry and colours for 1494 sources in the LRIS
field of view (Fig. \ref{lrisonfors}). Additional V-, R-, and I-band
images centered on \CL\ were obtained with the FORS1 (FOcal Reducer
and low dispersion Spectrograph; Appenzeller \& Rupprecht 1992) at the
ESO VLT and cover a field of view of 6\farcm8 $\times$ 6\farcm8
centered. Vega magnitudes are used throughout this paper.

\begin{table*}[t]
\begin{center}
\begin{tabular}{cccccccc}
\hline \hline
{{\normalsize Mask {\bf Name}}} & {{\normalsize Date}} & {{\normalsize Telescope}} & {{\normalsize Instrument}} & {{\normalsize Grism/Filter}} & {{\normalsize Exp. Time}} & {{\normalsize No. Exposures}} & {{\normalsize Mask}} \\
\hline
{\normalsize m6 } & {\normalsize Dec 1999 } & {\normalsize UT1 } & {\normalsize FORS1 } & {\normalsize 300I/none } & {\normalsize 1800 } & {\normalsize 2 } & {\normalsize MOS} \\
{\normalsize m1 } & {\normalsize Oct 2000 } & {\normalsize UT2 } & {\normalsize FORS2 } & {\normalsize 300I/none } & {\normalsize 1800 } & {\normalsize 4 } & {\normalsize MOS} \\
{\normalsize m2 } & {\normalsize Oct 2000 } & {\normalsize UT2 } & {\normalsize FORS2 } & {\normalsize 300I/none } & {\normalsize 1800 } & {\normalsize 4 } & {\normalsize MOS} \\
{\normalsize m3 } & {\normalsize Oct 2000 } & {\normalsize UT2 } & {\normalsize FORS2 } & {\normalsize 300I/none } & {\normalsize 1800 } & {\normalsize 4 } & {\normalsize MOS} \\
{\normalsize m4 } & {\normalsize Oct 2000 } & {\normalsize UT2 } & {\normalsize FORS2 } & {\normalsize 300I/none } & {\normalsize 1800 } & {\normalsize 8 } & {\normalsize MOS} \\
{\normalsize m5 } & {\normalsize Oct 2000 } & {\normalsize UT1 } & {\normalsize FORS1 } & {\normalsize 150I/none } & {\normalsize 1800 } & {\normalsize 1 } & {\normalsize MOS} \\
{\normalsize m7 } & {\normalsize Nov 2001 } & {\normalsize UT4 } & {\normalsize FORS2 } & {\normalsize 300I/none } & {\normalsize 1800 } & {\normalsize 6 } & {\normalsize MOS} \\
{\normalsize m8 } & {\normalsize Aug 2002 } & {\normalsize UT4 } & {\normalsize FORS2 } & {\normalsize 300I/none } & {\normalsize 2030 } & {\normalsize 4 } & {\normalsize MXU} \\
{\normalsize m9 } & {\normalsize Aug 2002 } & {\normalsize UT4 } & {\normalsize FORS2 } & {\normalsize 300I/none } & {\normalsize 2030 } & {\normalsize 6 } & {\normalsize MXU} \\
{\normalsize m10 } & {\normalsize Sep 2003 } & {\normalsize UT4 } & {\normalsize FORS2 } & {\normalsize 150I/none } & {\normalsize 1200 } & {\normalsize 13 } & {\normalsize MXU} \\
{\normalsize m11 } & {\normalsize Sep 2003 } & {\normalsize UT4 } & {\normalsize FORS2 } & {\normalsize 150I/none } & {\normalsize 1200 } & {\normalsize 1 } & {\normalsize MXU} \\
{\normalsize m11 } & {\normalsize Nov 2003 } & {\normalsize UT4 } & {\normalsize FORS2 } & {\normalsize 150I/none } & {\normalsize 1200 } & {\normalsize 3 } & {\normalsize MXU} \\
{\normalsize m11 } & {\normalsize Dec 2003 } & {\normalsize UT4 } & {\normalsize FORS2 } & {\normalsize 150I/none } & {\normalsize 1200 } & {\normalsize 10 } & {\normalsize MXU} \\
\hline \hline
\end{tabular}
\caption{A summary of the VLT spectroscopic survey on \CL. The first
column lists the mask name, the second column lists the observation
date, the third column lists the telescope and the fourth column lists
the instrument. After August 2002, all the observations were carried
out with the FORS2 CCD mosaic. The fifth column lists the grism and
the blocking filter used in the observation. The sixth column shows
the integration time per exposure and the seventh column shows the
number of exposures. The last column lists the mask type (MOS or MXU)
used.}
\label{tab_survey}
\end{center}
\end{table*}

\subsection{Spectroscopic observations and data reduction}\label{specred}

The spectroscopic observations of \CL\ were conducted in visitor and
service modes (VM and SM) from December 1999 to December 2003, as
summarized in Table \ref{tab_survey}. Bayesian photometric redshifts
(Ben\'{\i}tez 2000) were computed for each source using our BVRIJK$_s$
catalog. Candidates for spectroscopy were selected according to their
R-band magnitudes ($\mathrm{R} < 24$) and photometric redshifts ($ 0.7
< z_\mathrm{phot} < 0.95 $).  This photometric redshift interval was
chosen to match the dispersion in $z_\mathrm{phot}-z_\mathrm{spec}$ of
the $\sim 20$ redshifts obtained from the first few masks, which were
designed to target galaxies in the cluster red sequence (CRS).
Because of the larger FOV of FORS1 compared to the LRIS and SofI
fields of view (see Fig. \ref{lrisonfors}), no multi-band photometric
information was available for a number of sources outside the dashed
rectangle in Fig. \ref{lrisonfors}, and selections in this region were
carried out using only the FORS1 imaging.

At $z \sim 0.8$, spectral features such as the [OII]($\lambda$3727)
emission line, the $\mathrm{Ca II}$ H and K absorption lines and the
4000 \AA\ break, which are normally used to measure the redshift of a
galaxy, are between 6500 and 8500 \AA. An adequate coverage of this
range is provided with the 300I grism of FORS\footnote{When we wish to
refer to FORS1 and FORS2, we simply write FORS}. During poor seeing
conditions, we used the 150I grism, which provides additional coverage
below 5000 \AA, to observe relatively bright field galaxies with a
lower resolution.  No order-separation filters were used in order to
obtain extended coverage in the blue part of the spectrum.

A total of 11 masks for multi-object spectroscopy were prepared with
FIMS, the FORS mask preparation tool. Seven masks were designed for
use with the MOS (Multi Object Spectroscopy) mode and the other four
masks were designed for use with the MXU (Mask Exchange Unit)
mode. The MOS mode allows the positioning of up to 19 slits per mask,
each of which is 22\farcs5 long and formed by a pair of movable
blades.  In the MXU mode, each mask contains a larger number of slits,
which can be chosen to vary in length.  In order to remove fringes
accurately (see below), we chose a minimum slit-length of 10\arcsec\,
allowing us to place up to $\sim$ 40 slits per mask.

MOS observations were carried out with both FORS1 and FORS2 (see Table
\ref{tab_survey}), while the MXU mode was only available with
FORS2. Before March 2002, both FORS1 and FORS2 were equipped with
Tektronix CCD detectors with 2080 $\times$ 2048 pixels. During March
2002, FORS2 was upgraded with a mosaic of two 2k $\times$ 4k MIT CCDs
with a gap of 4\arcsec\ between them. As a result  the sensitivity in
the red increased by 30\%, allowing us to obtain about 30\% more
redshifts per mask.  Our data were taken in the standard resolution
mode, which provides a field of view of 6\farcm8 $\times$
6\farcm8. Slit widths of 1.0, 1.2, and 1.4 arcseconds were used in MOS
observations on both FORS1 and FORS2, which resulted in spectral
resolutions of $\sim$ 13 \AA, $\sim$ 16 \AA, and $\sim$ 18 \AA\
respectively for the 300I grism, and resolutions of $\sim$ 27 \AA,
$\sim$ 32 \AA, and $\sim$ 38 \AA\ respectively for the 150I grism. For
MXU observations, all slits were 1\arcsec\ wide, which corresponds to
resolutions of $\sim 28$ \AA\ and $\sim 13$ \AA\ for the 150I and 300I
grisms respectively. The main characteristics are summarized in Table
\ref{tab_instrument}.

\begin{table*}[t]
\begin{center}
\begin{tabular}{c||c|cc|cc}
\hline \hline
{{\small {\bf FORS Detector}}} & {{\small }} & {{\small {\bf Grism 
150I}}} & {{\small }} & {{\small {\bf Grism 300I}}} & {{\small }} \\
{{\small }} & {{\small Scale [arcsec/pix]}} & {{\small 
$\lambda_{range}$ [\AA]}} & {{\small Disp. [\AA/pix]}} & {{\small 
$\lambda_{range}$ [\AA]}} & {{\small Disp. [\AA/pix]}} \\
\hline
\hline
{\small Tektronix CCD} & {\small 0.20} & {\small 4000-10000} & {\small 
5.52} & {\small 6000-10000} & {\small 2.59} \\
\hline
{\small 2k$\times$4k MIT CCD} & {\small 0.25} & {\small 4000-10000} & 
{\small 6.90} & {\small 6000-10000} & {\small 3.24} \\
\hline \hline
\end{tabular}
\caption{The characteristics of spectroscopic modes used in the
survey. For more information, see the FORS User Manual issue 2.6 at
http://www.eso.org/instruments/fors1/userman/index.html .}
\label{tab_instrument}
\end{center}
\end{table*}

During the MXU observations, the targets were dithered along the
direction of the slits in order to allow for the removal of fringes in
the spectra. This fringing pattern may be quite strong at wavelengths
beyond 7000 \AA, complicating the identification of spectral features
in  the red. Indeed, our first MOS observations with the Tektronix
CCD, carried out with no dithering, showed the importance of
offsetting the telescope between exposures. With the new FORS2 MIT
CCD mosaic; however, the fringe amplitude was determined to be
$\simless$ 5\%. Nevertheless, fringe removal was still needed to reach
the Poisson noise limit. Both VM and SM observations were carried out
under seeing conditions of $<$ 0\farcs8.

To assist in data reduction, we developed a specially dedicated
IRAF\footnote{Image Reduction and Analysis Facility. See
http://iraf.noao.edu/iraf-homepage.html} pipeline, which included
shell scripts.  This pipeline was optimised to reduce data taken with
the FORS2 MIT CCD mosaic, controlling the separate reduction of each
chip in an interactive way and merging the two reduced frames to
produce a final single image.   Standard IRAF tasks were implemented
to perform bias and flat corrections, and once the single frames
corresponding to a given mask were obtained, the slitlets for each
exposure were grouped according to slit number.  Single slit spectra
were then background subtracted and (when dithered exposures were
available) fringe corrected  using an algorithm similar to  that
implemented in BOGUS~\footnote{BOGUS is available on line at
http://astron.berkeley.edu/$\sim$dan/homepage/bogus.html}, which was
developed by D. Stern, A.J. Bunker and S.A. Stanford. The sky was
subtracted using the standard IRAF routines, and the fringe-corrected
and sky-subtracted frames were registered and combined. Extraction,
wavelength and flux calibration were carried out in a standard
manner. Signal-to-noise ratios varied from values lower than 1, in
cases where no continuum but only emission lines were detected, up to
values of 50 or more per pixel at about 4000 \AA\ in the rest
frame. At a redshift of $z=0.84$, the H$_{\delta}$ ($\lambda$4101.7)
absorption feature falls at about 53 \AA\ from the atmospheric A-band
at about 7600 \AA, which was removed with the IRAF task TELLURIC.

Finally, the redshifts were measured using the cross-correlation
technique (e.g., Tonry \& Davis 1979) implemented in the IRAF task
XCSAO (Kurtz et al. 1992). The observed spectra were correlated with
templates from Kinney et al. (1996), and the typical errors computed
by XCSAO were $\delta z \sim 3 \times 10^{-4}$. Multiple observations
of some targets allowed us to estimate total (random+systematic)
errors, yielding typical values of $\delta z \sim 8 \times 10^{-4}$.

\begin{center}
\begin{figure}[t]
\centerline{\psfig{file=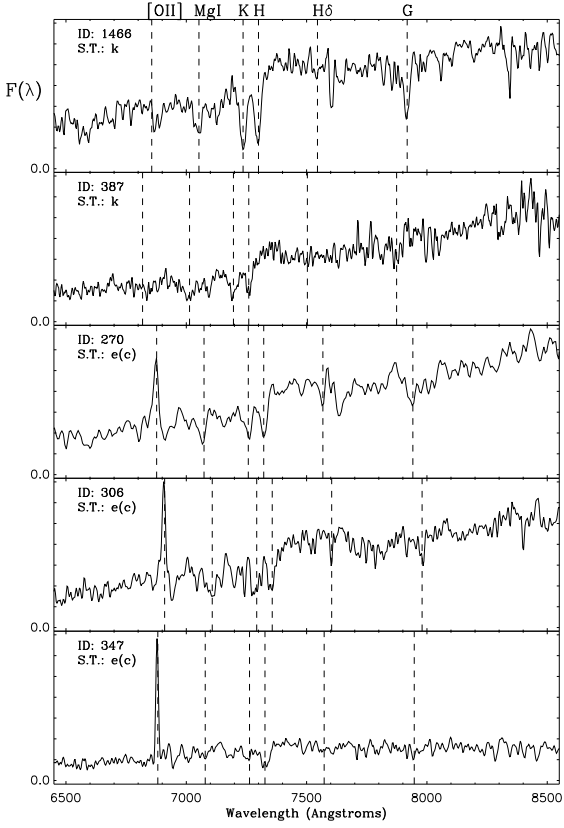,width=8cm}}
\caption{Spectra of five cluster members. The spectra are arranged top
to bottom from early type galaxies to late type galaxies. The most
common spectral features are indicated by the dashed lines. From left
to right they correspond to [OII]($\lambda$3727), MgI($\lambda$3834),
$\mathrm{Ca II}$ K($\lambda$3934), $\mathrm{Ca II}$ H($\lambda$3969),
H$_{\delta}$($\lambda$4102) and the G-band ($\lambda$4304). The flux
axis is in arbitrary units. The spectroscopic type (S.T.) as defined
in Dressler et al. (1999) and the candidate number (ID) are also
indicated.}
\label{memb_spectra}
\end{figure}
\end{center}

\begin{center}
\begin{figure}[t]
\centerline{\psfig{file=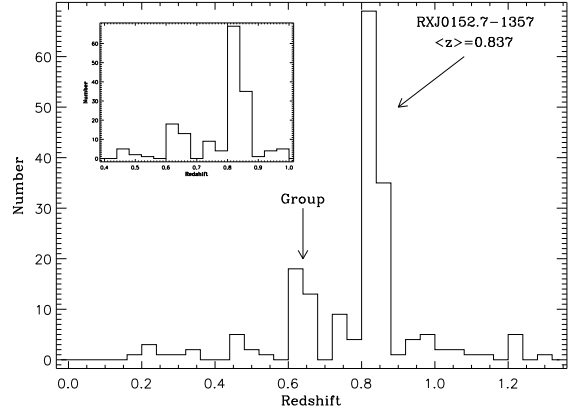,width=7cm,angle=-90}}
\caption{Redshift distribution of 187 galaxies with secure
redshifts. The bin size is $\Delta z=0.04$. The main peak in the
distribution corresponds to \CL\ with 102 galaxies within the range
$0.81 < z < 0.87$. A secondary peak is observed, indicating the
existence of a group of galaxies at $z \sim 0.64$. The inset shows a
zoom of the redshift distribution in the $0.4 < z < 1.0$ range.}
\label{redshift_histo}
\end{figure}
\end{center}

\begin{center}
\begin{figure*}[t]
\centerline{\psfig{file=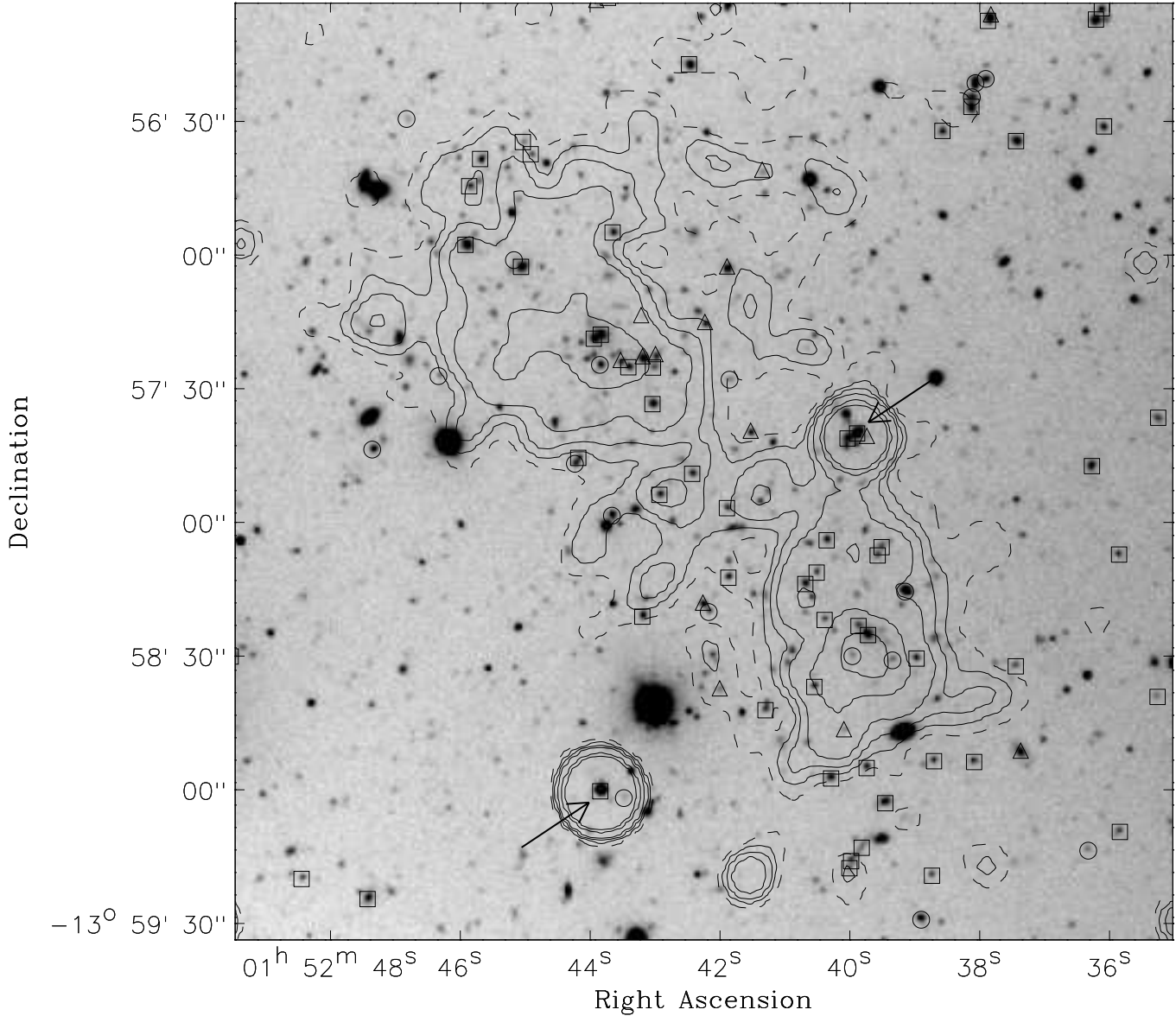,width=18cm}}
\caption{The distribution of confirmed cluster members in the central
region of \CL. The field of view in this figure is 3\farcm5 $\times$
3\farcm5 ( 1.6 Mpc $\times$ 1.6 Mpc at $z\sim0.84$). North is up and
East is to the left. The squares mark the location of cluster members
and the contours correspond to X-ray emission (3, 5, 7, 10, 20 and 30
$\sigma$ above the background in the [0.5-2] keV band) measured with
Chandra. The galaxy distribution follows that of the hot intra-cluster
gas in the high-density regions. The arrows indicate the location of
the two AGNs that are cluster members. Also shown are field galaxies
(circles) and galaxies belonging to the foreground group (triangles)
at $z \sim 0.64$.}
\label{gal_x_dist}
\end{figure*}
\end{center}

\subsection{The spectroscopic catalog}

A total of 262 objects were spectroscopically observed with FORS1 and
FORS2 between December 1999 and December 2003. We were able to obtain
redshifts for 227 (87\%) objects, out of which 187 (72\%) are secure
and 40 (15\%) are less secure due to low signal-to-noise.

We classify galaxies as cluster members if they have a secure redshift
within $0.81 < z < 0.87$. In this interval, galaxies are within the
$\pm 3\sigma_v$ region around the cluster median velocity and the
cluster structure appears well isolated in redshift space. The
velocity dispersion of \CL, $\sigma_v$, is calculated in section \S
\ref{gal_dist}. With this criterion, 102 objects are classified as
cluster members. The corresponding spectroscopic catalog is presented
in Table 4. Spectroscopic redshifts and error bars are those computed
by XCSAO (Kurtz et al. 1992; see section \S\ref{specred}). The most
prominent spectral features of each spectrum are indicated in the
seventh column of the catalog (Table 4). An emission line flag was
given to each object. A value of 0 corresponds to galaxies without
observed emission lines, a value of 1 corresponds to galaxies showing
any emission line. Two galaxies with broad emission lines (AGNs) were
assigned a value of 2. As an example of the data obtained during our
survey, the spectra of 5 cluster members are shown in
Fig. \ref{memb_spectra}. From top to bottom, the spectra are arranged
from early type galaxies to late type galaxies. The galaxy
spectroscopic type as defined in Dressler et al. (1999) is also
given. The most common spectral features are indicated by dashed
lines, as explained in the figure caption.

The number of confirmed cluster members obtained in our spectroscopic
survey is comparable to other surveys of distant clusters as, e.g.,
MS1054-03 (Donahue et al. 1998; Tran et al. 1999; van Dokkum et
al. 1999; van Dokkum et al. 2000) and the supercluster Cl 1604+4304 at
$z=0.90$ (Gal \& Lubin 2004). In addition to the number of secure
members, we have 15 objects at the cluster redshift with less secure
redshifts. In Table 5 we also present the coordinates and redshifts of
the non-cluster members.

The ground-based photometric catalog itself was used to estimate the
success rate of our spectroscopic survey. Since the imaging data were
obtained with LRIS, we restrict the analysis to the 4\farcm9 $\times$
6\farcm54 field of view of LRIS. We computed the ratio of the number
of object with spectroscopic redshifts to the number of objects in the
photometric catalogue that were targeted for spectroscopy as a
function of R magnitude. The data were binned in $\Delta R=0.5$ mag
intervals. The ratio is observed to be nearly constant between the
interval $20 < R < 24$ with a mean value of 0.89 and a dispersion of
0.07.

\section{Analysis}

\subsection{The survey redshift distribution}\label{redshift_survey}

Fig. \ref{redshift_histo} shows the distribution in redshift space of
the 202 galaxies for which a secure estimate of the redshift was
obtained. The bin size of the histogram was fixed to $\Delta
z=0.04$. \CL\ clearly appears as the most significant structure in the
distribution, with a median redshift of $z=0.837$ (3$\sigma$-clipped
mean of $\bar{z}=0.8376 \pm 0.0010$) and 102 galaxies between $z=0.81$
and $z=0.87$. All these galaxies are located in the 6\farcm8 $\times$
6\farcm8 region covered by FORS1, which is equivalent to an area of
3.1 Mpc $\times$ 3.1 Mpc at the cluster redshift. A secondary peak in
the histogram is clearly seen at $z \sim 0.64$, where 31 galaxies have
been confirmed in the range $0.60 < z < 0.68$. The spatial
distribution of this group of galaxies is quite elongated in the
North-South direction, extending across the field of view and passing
in front of the cluster as shown in Fig. \ref{gal_x_dist}. Its
morphology suggests that this structure may be a filament
perpendicular to the line of sight, with a projected overdensity of
galaxies on top of the northern substructure of \CL\ (see section
\S\ref{gal_dist}). The presence of this foreground structure should be
taken into account when estimating the mass of the cluster through the
weak lensing technique, which is sensitive to the distribution of
matter along the line of sight. No X-ray emission seems to be
associated to this structure, indicating that the galaxies in this
group do not have a hot gas component.

\begin{center}
\begin{figure}[t]
\centerline{\psfig{file=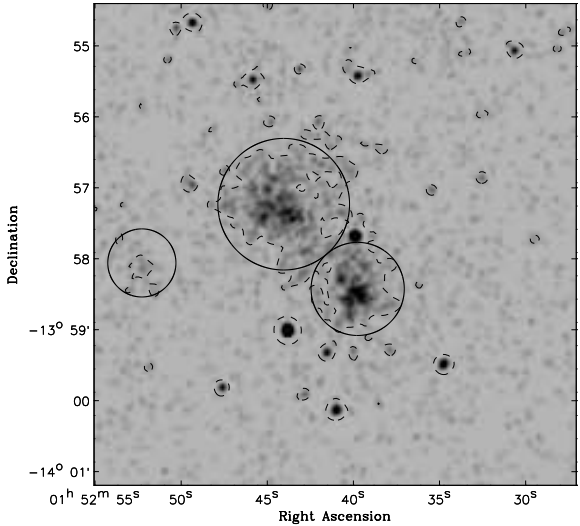,width=8cm}}
\caption{Chandra image of \CL. The field of view is 6\farcm8 $\times$
6\farcm8 and is centered on the cluster, North is up and East is to
the left. The dashed contour corresponds to the 3$\sigma$ level above
the background. The circles define the three main regions (northern,
southern and eastern clumps) where the extended X-ray emission is
greater than 3$\sigma$ times the background.}
\label{x_3sig}
\end{figure}
\end{center}

\subsection{Galaxy distribution and substructure in \CL}\label{gal_dist}

The projected distribution of cluster members in the central part of
\CL\ is presented in Fig. \ref{gal_x_dist}, where cluster members are
marked with squares. The X-ray emission, field galaxies (circles) and
galaxies belonging to the group at $z \sim 0.64$ (triangles) are also
shown. Following the high-density regions as traced by the hot
intracluster gas, the cluster members are grouped into two main
clumps, one to the north-east of the center of the image (hereafter
called northern clump) and the other one to the south-west (hereafter
called southern clump). These two clumps are separated by $\sim$
1\farcm6, which corresponds to $\sim$ 730 kpc at z $\sim$ 0.84, and
coincide with the two main peaks observed in the X-ray. Another
accumulation of cluster members is observed about 2\farcm4 ($\sim$ 1
Mpc at the cluster redshift) to the East of the central double
structure. It is interesting to note that this third clump (hereafter
called the eastern clump) also coincides with a region where diffuse
X-ray emission is detected with a significance greater than 3$\sigma$
over the background. These findings clearly support the picture that
\CL\ is a dynamically young cluster still in formation, with two and
possibly three substructures in an on-going merging phase. Indeed,
Maughan et al. (2003) show some evidence for an adiabatic collision in
progress between the northern and southern clumps. Additionally, there
is an overdensity of cluster members (about 11 galaxies) located to
the north-west of the northern clump, although there is no diffuse
X-ray emission associated with this clump.

By studying the weak lensing signal around the cluster from HST/ACS
data, Jee et al. (2004) find a projected distribution of dark matter
(DM) that is in close agreement with both the observed 2-dimensional
distribution of galaxies and the ICM in \CL. The DM map shows a central
elongated structure with a strong concentration that coincides with
the northern clump. Moreover, the eastern clump is also observed in
the weak lensing map, confirming its existence.

\begin{center}
\begin{figure*}[t]
\centerline{\psfig{file=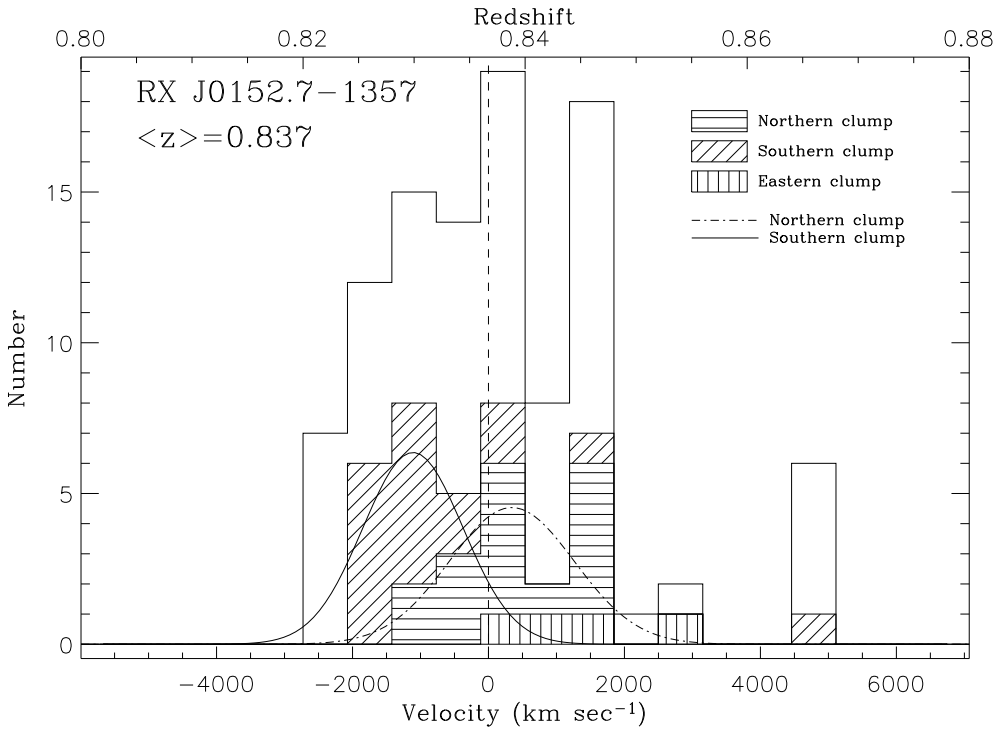,width=10cm,angle=90}}
\caption{A velocity histogram of the galaxies in \CL. The unshaded
histogram shows the velocity distribution of 102 cluster members,
while the shaded histograms correspond to the velocity distributions
in the three circular regions encompassing the main substructures. For
the sake of clarity, shaded histograms have been stacked instead of
overlayed. The two {G}aussians represent the distribution of
galaxies in the northern and southern clumps. Their location (median
redshift) and scale (velocity dispersion) are those given in the
text.}
\label{cluster_histo}
\end{figure*}
\end{center}

The observed substructure in the projected galaxy distribution was
also studied in velocity space. We concentrated on substructures that
are associated with the diffuse X-ray emission, which were analysed as
follows. We separated galaxies located in the northern clump from
those located in the southern clump by comparing the Chandra data with
the VLT data. Fig. \ref{x_3sig} shows a 6\farcm8 $\times$ 6\farcm8
portion of the Chandra image centered on the cluster. The 3-$\sigma$
level above the background is indicated by the dashed contour. Three
circular regions were defined, each centered on one of the three X-ray
peaks. The radius of each region was adjusted in order to encompass
most of the X-ray emission above the 3$\sigma$ level. The radii for
the northern, southern and eastern clumps are 0\farcm93 ($\sim$430
kpc), 0\farcm66 ($\sim$300 kpc), and 0\farcm48 ($\sim$220 kpc)
respectively. Fig. \ref{lrisonfors} shows the 3$\sigma$ X-ray
iso-contours (dashed curve) and the circular regions overlaid on the
FORS1 image. A total of 16 cluster members are located within the
circle corresponding to the northern clump, 18 are within the southern
clump and 4 are found within the eastern clump. Note that the
association of a galaxy to any given clump is based only on the
projected distance. The median redshifts of these galaxy clumps are
$z_N=0.8388$, $z_S=0.8299$ and $z_E=0.8473$ respectively. The
corresponding 3$\sigma$-clipped mean redshifts and their standard
deviations are: $\bar{z}_N=0.8390 \pm 0.0013$, $\bar{z}_S=0.8299 \pm
0.0009$ and $\bar{z}_E=0.8458 \pm 0.0027$.

The velocity histogram in the cluster rest frame is presented in
Fig. \ref{cluster_histo} (the vertical dashed line corresponds to
z=0.837). The unshaded histogram shows the distribution of the 102
cluster members, while the shaded histograms correspond to the
distributions in the three circular regions defined above and
indicated in Fig. \ref{lrisonfors}. For the sake of clarity, the
shaded histograms have been stacked instead of overlayed. A close
inspection of the velocity histogram shows that the kinematic behavior
of the galaxies in the two main clumps are different. Indeed, a large
fraction of the galaxies in the southern clump have negative {\it
peculiar} velocities, despite some outliers moving at velocities
greater than 2000 km s$^{-1}$. The northern clump instead has a mean
velocity close to the median of the overall velocity distribution. In
addition to this, the four galaxies in the eastern clump are observed
to have positive {\it peculiar} radial velocities. Based on Tukey's
biweight estimator, we compute the velocity dispersion for the two
main galaxy clumps, obtaining $\sigma_N=919 \pm 168$ km s$^{-1}$ and
$\sigma_S=737 \pm 126$ km s$^{-1}$. These numbers should be taken with
caution, because of the small number of redshifts, the simple
analysis, and because the distribution of galaxies may be less
virialised than the gas, as the ICM is expected to reach an
equilibrium state in a shorter time scale compared to galaxies and
dark matter particles. Nevertheless, these values are in agreement
with what is expected from X-ray temperatures (Wu, Fang \& Xu 1998),
indicating that the galaxies in these two sub-clumps are probably
virialised. Two {G}aussians with median redshifts $z_N$ and $z_S$ and
velocity dispersions $\sigma_N$ and $\sigma_S$ are also shown in
Fig. \ref{cluster_histo} to represent the distribution of galaxies in
the northern and southern clumps respectively. A least-squares fit of
a Gaussian to all the cluster members was performed, yielding a
velocity dispersion of is $\sigma_v=1546 \pm 208$ km s$^{-1}$. A more
robust estimate by using the Tukey's biweight estimator delivers a
value of $\sigma_v=1632 \pm 115$ km s$^{-1}$. Assuming a virialised
state, a velocity dispersion of $\sigma_v=1600$ km s$^{-1}$
corresponds to an X-ray temperature of $T_\mathrm{X}=16$ keV, which is
significantly higher than the observed value, therefore the overall
structure of the cluster is not in dynamical equilibrium.

We anticipate that the results obtained in this section are in good
agreement with a more specific kinematic analysis presented in Girardi
et al. (in preparation). The latter includes the use of robust
statistical indicators (e.g., Beers et al. 1990) to determine velocity
dispersions and a formal identification of substructures (e.g.,
Dressler \& Schectman 1988; Mercurio et al. 2003a) in velocity
space. We refer the reader to Girardi et al. for a more complete
analysis of the kinematical structure of \CL\ and its mass estimate.

\subsection{Mass and luminosity estimates}

A simple estimation of the mass of the northern and southern
subclusters can be obtained from (see e.g. Longair 1998):

\begin{equation}
M (R)=3 \frac{\sigma^2_v R}{G} \ , 
\label{virmass}
\end{equation}

\noindent
where G is the gravitational constant and $\sigma_v$ is the velocity
dispersion along the line of sight obtained above. The velocity
dispersions $\sigma_N$ and $\sigma_S$ (see \S\ref{gal_dist}) for the
northern and southern clumps respectively have been used together with
the radii defined for the same substructures, $r_N=430$ kpc and
$r_S=300$ kpc (see \S\ref{gal_dist} and Fig. \ref{lrisonfors}), to
estimate their masses via Eq. \ref{virmass}. We find $M_N=(2.5 \pm
0.9) \times 10^{14} M_{\odot}$ and $M_S=(1.1 \pm 0.4) \times 10^{14}
M_{\odot}$ for the northern and southern subclusters respectively. The
sum of these two values gives $M=(3.6 \pm 1.0) \times 10^{14}
M_{\odot}$. These values are in good agreement with the results
obtained by Huo et al. (2004) based on a joint analysis of Chandra and
Keck data. Our mass estimates for the northern and southern
subclusters are also consistent with the mass estimates within
$r_{500}$ (the radius within which the mean density is 500 times the
critical density of the universe at the given redshift) presented by
Ettori et al. (2004) for the same substructures.

It is not simple to make a direct and clean comparison of our mass
values to those given in the weak lensing analysis of Jee et
al. (2004) and the X-ray study of Maughan et al. (2003).  The mass
clumps C and F in Jee et al. (2004) correspond to our northern and
southern subclusters respectively. Our eastern clump corresponds to
mass clump A in the weak lensing map. The mass values in Table 2 of
Jee et al. (2004) are given within an aperture of 20\arcsec\
($\sim$152 kpc) in radius. Within the same aperture, we obtain about
68\% less spectroscopically confirmed galaxies in the main central
clumps than those reported above (see \S\ref{gal_dist}), for which our
velocity dispersion and therefore mass estimates would not be
reliable. However we note that the mass value of $(2.1 \pm 0.3) \times
10^{14} M_{\odot}$ obtained by Jee et al. (2004) within an aperture of
50\arcsec\ ($\sim$380 kpc) in radius centered on the northern clump is
in very good agreement with our mass estimate given above for the same
substructure. Apertures with radii larger than $r_N$ and $r_S$ would
not be adequate to measure the mass of the northern and southern
clumps respectively since they would also include galaxies outside the
selected substructure, affecting thus its mass estimate. Therefore we
do not compare masses to those in Jee et al. (2004) and Maughan et
al. (2003) computed within 1 Mpc and 1.4 Mpc radius
apertures. However, we note that the value of $2.4^{+0.4}_{-0.3}
\times 10^{14} M_{\odot}$ in Maughan et al. (2003), obtained from
combining the masses of the two subclusters within apertures of about
50\arcsec\ ($\sim$380 kpc) are in good agreement, within the
uncertainties, with our combined mass of the northern and southern
subclusters given above.

Since the overall galaxy distribution in \CL\ is not in virial
equilibrium, our velocity dispersion value from the 102
spectroscopically confirmed cluster members leads us to overestimate
the true total mass of the cluster. Considering an aperture of $R=1$
Mpc our value of $\sigma_v=1632 \pm 115$ km s$^{-1}$ yields a
$M_{CL}=(1.9 \pm 0.3) \times 10^{15} M_{\odot}$, about 4 times larger
than the value obtained from the weak lensing analysis (Jee et
al. 2004) within the same aperture. From Eq. \ref{virmass}, the mass
within the virial radius of R=1.4 Mpc (see Maughan et al. 2003)
associated to the same value of $\sigma_v$ turns out to be $M_{CL} =
(2.6 \pm 0.4) \times 10^{15} M_{\odot}$, about within a factor of two
of the combined total mass of the two subclusters extrapolated to the
virial radius in Maughan et al. (2003) paper.

Our ground based I-band photometry was used to estimate the total
luminosity in the B-band, $L_B$, from the cluster galaxies within a
given aperture. The large number of spectroscopically confirmed
cluster members, which are likely to be the bright ones and thus
dominate the total optical cluster light, allow us to make a fair
estimate of $L_B$. By using the galaxy templates in Ben\'{i}tez et
al. (2004) and following the procedure described in Cross et
al. (2004) to compute rest-frame B-band magnitudes, $B_{rest}$, we
obtain the transformation from our I-band photometry to the $B_{rest}$
magnitudes to be $B_{rest}= I - dm - k_{BI}$, where $dm$ is the
distance modulus of the cluster and $k_{BI}$ is the general
$k$-correction term computed from the template galaxy spectra. This
depends on the B- and I-band filters and redshift. At the cluster
redshift of z = 0.837 we obtain values of $dm$ = 43.62 and $k_{BI}$ =
-1.35, the latter being only weakly dependent on galaxy type. The
total B-band luminosity is then computed as:

\begin{equation}
\frac{L_B}{L_{B,\odot}}=\sum_j 10^{0.4\{ M_{B,\odot} - B_{rest, j}\}} \ ,
\label{ltot}
\end{equation}

\noindent 
where $M_{B,\odot}=5.48$ is the absolute B magnitude of the sun,
$L_{B,\odot}$ is the solar luminosity in the B band and the sum is
extended to all the objects within the given aperture. In order to
compare our results with those from the weak lensing analysis, we
compute $L_B$ within an aperture of 1 Mpc centered on the northern
clump. From Eq. \ref{ltot} we obtain $L_B=4.8 \times 10^{12}
L_{B,\odot}$, which is in excellent agreement with the value of
$L_B=5.2 \times 10^{12} L_{B,\odot}$ given in Jee et
al. (2004). However, since we only considered spectroscopically
confirmed members, our value of $L_B$ may be underestimated.

As was the case in comparing mass estimates, a direct and clean
comparison of the mass-to-light ratios obtained in this paper for the
northern and southern clumps to the values given in the weak lensing
paper (Jee et al. 2004) cannot be done. Within our apertures of
$r_N=430$ kpc and $r_S=300$ kpc in radius we obtain mass-to-light
ratios of $(246 \pm 89) M_{\odot}/L_{B,\odot}$ and $(152 \pm 55)
M_{\odot}/L_{B,\odot}$ for the northern and southern subclusters
respectively. As a reference, the $M/L_{B}$ value obtained by Jee et
al. (2004) within a 1 Mpc aperture centered on the northern clump is
$(92 \pm 7) M_{\odot}/L_{B,\odot}$, significantly lower than our
values for any of the main subclusters. However, since we only count
spectroscopically confirmed cluster members, it is possible that we
are missing some of the B-band light from unconfirmed cluster galaxies
and that we are likely overestimating the $M/L_B$ ratios. Our value of
$M/L_B$ for the northern clump is about 1.8 times higher than the
expected one from the mass-to-light ratio profile presented in Fig. 19
of Jee et al. (2004).

\begin{center}
\begin{figure}[t]
\centerline{\psfig{file=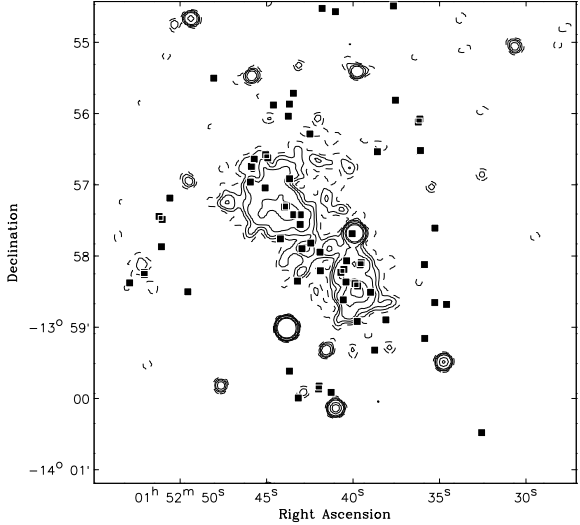,width=8cm}}
\caption{The projected distribution of cluster members without
emission lines (filled squares). X-ray iso-contours are also shown.}
\label{noel_membs}
\end{figure}
\end{center}

\begin{center}
\begin{figure}[t]
\centerline{\psfig{file=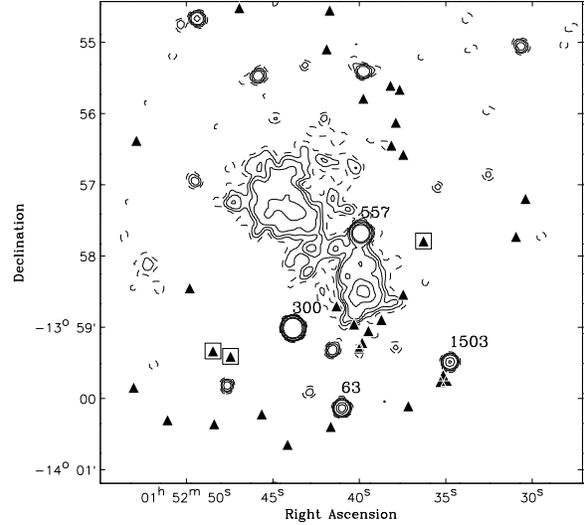,width=8cm}}
\caption{The projected distribution of cluster members with emission
lines (filled triangles). X-ray iso-contours are also shown. Open
squares indicate the location of red star-forming galaxies. The
spectroscopically observed AGN are labeled with their corresponding
IDs and their redshifts are given in Table \ref{tab_agns}.}
\label{el_membs}
\end{figure}
\end{center}

\begin{center}
\begin{figure}[t]
\centerline{\psfig{file=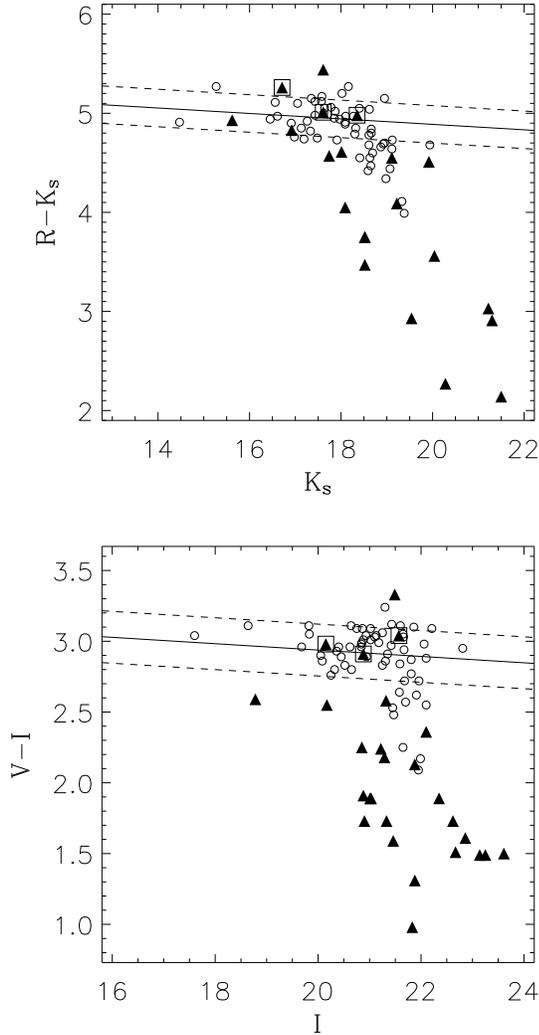,width=16.7cm,angle=-90}}
\caption{Colour-Magnitude diagrams of cluster members within the LRIS
field of view. Open circles are passive galaxies and solid triangles
are star-forming galaxies. The solid line corresponds to a linear fit
to the data within the cluster red sequence (CRS) defined as the [$14
< K_s < 18.5$, $4.5 < R-K_s < 6$] region in the upper panel and the [$
16 < I < 22$, $2.3 < V-I < 3.5$] region in the lower panel. Dashed
lines indicate the 1-$\sigma$ level above and below the linear fit
($\sigma=0.19 \pm 0.02$ in upper panel and $\sigma=0.18 \pm 0.02$ in
lower panel). The large scatter is in part due to the shallowness of
our ground based photometry. The reddest galaxies with narrow emission
lines in the Colour-Colour diagrams of Fig. \ref{colcol} are indicated
by open squares.}
\label{rcs}
\end{figure}
\end{center}

\begin{center}
\begin{figure}[t]
\centerline{\psfig{file=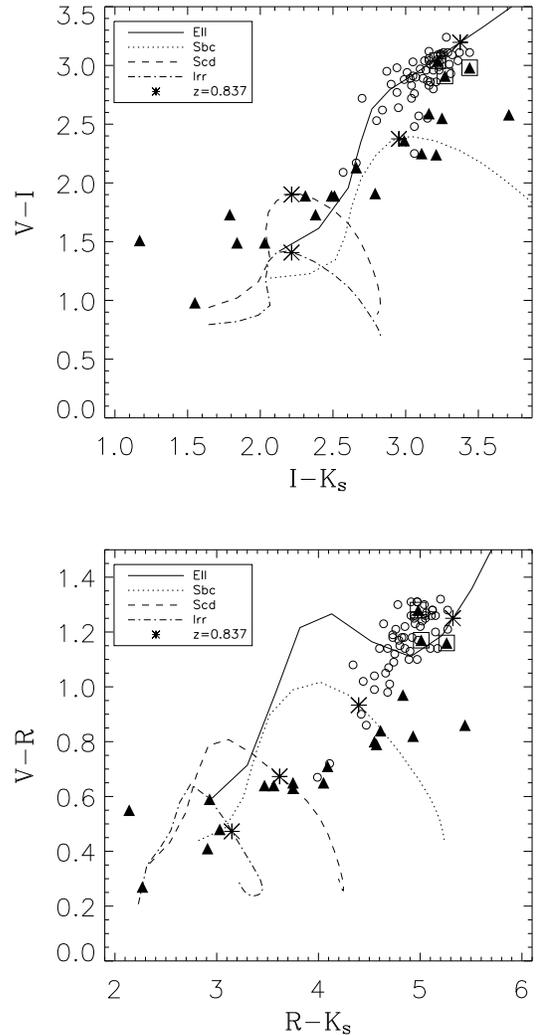,width=16.7cm,angle=-90}}
\caption{Colour-Colour diagrams of cluster members within the LRIS
field of view. Open circles are passive galaxies and solid triangles
are star-forming galaxies. The reddest galaxies with narrow emission
lines are indicated by open squares. Colour-colour tracks based on
model spectra of Ell, Sbc, Scd and Irr galaxies without evolution are
also shown. Expected colours for the different galaxy types at the
cluster redshift are indicated with stars.}
\label{colcol}
\end{figure}
\end{center}

\subsection{Galaxy populations and substructure}

We use the spectra to divide cluster members into two subsets: those
with emission lines (star-forming galaxies) and those without clear
emission lines (passive galaxies). Figures \ref{noel_membs} and
\ref{el_membs} show the projected distribution of these subsets. In
our spectroscopic sample of cluster members, 34 (33\%) cluster
members, excluding the 2 AGN, have clear emission line features, such
as [OII] ($\lambda$3727) or [OIII] ($\lambda$5007), indicating the
existence of on-going star-formation.

The three high-density regions in the cluster mass distribution, as
traced by the hot X-ray gas (see Fig. \ref{x_3sig}), are populated by
passive galaxies exclusively. All the star-forming galaxies noticeably
avoid these regions.  A few galaxies with narrow emission lines are
observed close to the southern edge of the southern clump (see
Fig. \ref{el_membs}). These galaxies are about 30\arcsec\ ($\sim$ 230
kpc) from the centroid of the southern X-ray substructure, which
corresponds to a region where the local ICM density is about one-third
of the central density of the southern clump, as derived from the
X-ray surface brightness profile.

In terms of spectral classification, passive galaxies are split into
two main groups: k type galaxies (see Dressler et al. 1999; Poggianti
et al. 1999) and post-starburst galaxies (Dressler \& Gunn 1983), also
called ``E+A'' galaxies. Based on the strength of the
H$_{\delta}$($\lambda$4102) line, post-starburst galaxies can be
further divided into k + a and a + k type galaxies. (Franx 1993;
Dressler et al. 1999; Poggianti et al. 1999). Northern, southern and
eastern clumps are mostly populated by k type galaxies and a few k + a
and a + k galaxies are also seen in these clumps. The classification
as post-starburst galaxies of two out of ten cluster members may be
uncertain due to the possible alteration of the H$_{\delta}$
equivalent width by the telluric correction.

Most passive galaxies have red colours ($R-K_s \simeq 5$) and most of
them populate the cluster red sequence (CRS) in the colour-magnitude
diagram (CMD). In Fig. \ref{rcs}, we show the distribution of cluster
members that have ground based LRIS and SofI photometry in two
CMDs. Open circles are passive galaxies and solid triangles are
star-forming galaxies. The solid lines correspond to linear fits
($R-K_\mathrm{s} = -0.027 \ K_\mathrm{s} + 5.44$ in upper panel and
$V-I = -0.022 \ I + 3.39$ in lower panel) to the data within the
regions defined in the caption of Fig. \ref{rcs}. Dashed lines
indicate the 1-$\sigma$ level above and below the fits. If we compare
the CRS of the $R-K_\mathrm{s}$ vs $K_\mathrm{s}$ diagram to the
models computed by Kodama \& Arimoto (1997), we infer that the colours
of the passive galaxies in the CRS are consistent with an early
formation epoch ($z \simgreat 2$). However, we note the presence of a
tail of fainter ($K_s \simgreat 18.5$) passive galaxies extending to
bluer ($4.0 \simless R-K_s \simless 5.7$) colours, off the
CRS. Passive galaxies are also observed outside the high-density
peaks, however the outskirts of \CL\ are dominated by star-forming
galaxies.

The segregation between star-forming and passive galaxies in \CL\ is
in agreement with a picture in which the star formation activity in
galaxies is suppressed during the infall of these galaxies into the
highest density zones. An analysis of the physical properties of the
star-forming galaxy population in \CL\ based on the combination of VLT
spectroscopic and HST/ACS imaging data, together with implications for
cluster galaxy evolution, will be presented in Homeier et
al. 2004.

In Fig. \ref{colcol}, we show the distribution of cluster members in
two colour-colour diagrams. The symbols are the same as those in
Fig. \ref{rcs}. For reference, we also show the tracks as a function
of redshift for an elliptical (Ell) and three late-type (Sbc, Scd and
Irr) galaxies using the template library of Coleman, Wu, \& Weedman
(1980), whose spectral energy distributions (SEDs) have been extended
to the near-IR and far-UV using Bruzual \& Charlot (1993, with recent
updates) models. The tracks include k-corrections only. The expected
colours for the different galaxy types at the cluster redshift are
indicated with stars. Passive galaxies are located in the upper right
part of both colour-colour diagrams (see Fig. \ref{colcol}). The
observed offset between the locus of these galaxies and their expected
colours from the models is consistent with passive evolution of
early-type galaxies formed at $z \sim 1.9$.

Finally, we note that the brightest cluster member of the southern
clump (ID=387) is redder than the brightest cluster member of the
northern clump (ID=701). The differences in the colours are
$\Delta(R-K_\mathrm{s})=0.17$ and $\Delta(V-I)=0.15$ with errors of
$\delta(R-K_\mathrm{s})=0.015$ and $\delta(V-I)=0.012$
respectively. These differences in colour get bigger if we compare
ID=387 with the pair of bright galaxies (ID=1466 and ID=1467) located
in the core of the northern clump. Galaxy ID=387 turns out to be $\sim
0.25$ magnitudes redder in $R-K_\mathrm{s}$ and $\sim 0.24$ magnitudes
redder in $V-I$ than both galaxies at the core of the northern
subcluster. Assuming that these galaxies have similar metallicities,
such a difference in colour may be an indication of a difference in
stellar ages. However, more accurate photometry, such as, e.g., ACS
photometry, would be needed to confirm this result (Blakeslee et al.,
in preparation).

\begin{center}
\begin{figure}[t]
\centerline{\psfig{file=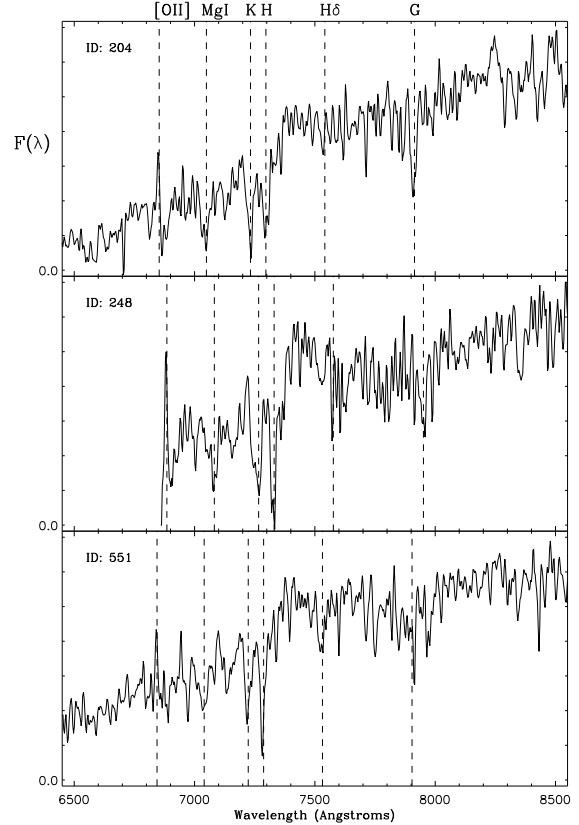,width=8cm}}
\caption{Spectra of red star-forming galaxies. The ID number of the
shown spectra are, from top to bottom: 204, 248 and 551. Important
spectral features are indicated with dashed lines (for a comparison, see
Fig. \ref{memb_spectra}). The flux is in arbitrary units. Due
to the location of galaxy 248 in the mask, there is no data for
wavelengths blueward of the [OII] ($\lambda$3727) line.}
\label{rsf_spectra}
\end{figure}
\end{center}

\subsubsection{Red star-forming galaxies}

Although most of the star-forming galaxies are seen to populate the
lower right region of the CMD, there are a few that are located in the
RCS and in the locus occupied by early-type galaxies in the
colour-colour diagrams in Fig. \ref{colcol}.  The three reddest
galaxies with emission features are indicated by open squares (see
Figs. \ref{rcs} and \ref{colcol}) and represent about 5\% (3 out of
64) of the galaxies within the CRS. Their red colours ($R-K_s \simeq
5$) suggest that these galaxies are dominated by an old stellar
population, although dust extinction can significantly affect the
observed colours. The spectra of these galaxies are shown in
Fig. \ref{rsf_spectra}. In addition to the [OII] ($\lambda$3727)
emission line observed in their spectra, 2 of these galaxies show a
prominent $H_{\delta}$ ($\lambda$4102) feature in absorption,
indicating the existence of a coeval young stellar population ($\sim$
1 Gyr). In the case of galaxy ID=248, $H_{\delta}$ ($\lambda$4102)
lies at about 23 \AA\ from the subtracted A-band feature which may
have introduced an uncertainty of $\sim$20\% in its equivalent width
after the A-band subtraction procedure.

A measurement of the $\mathrm{Ca II}$ H + H$\epsilon$ to $\mathrm{Ca
II}$ K ratio (Rose 1984, 1985) indicates that galaxies ID=248 and
ID=551 have a significant post-starburst component. In the case of
galaxy ID=204, the $\mathrm{Ca II}$ H + H$\epsilon$ to $\mathrm{Ca
II}$ K ratio turns out to be inconsistently larger than the observed
index values for different MK spectral types (Rose 1985). As observed
in the 2-d spectrum of this object, the inconsistency may be due to a
sky line effecting the $\mathrm{Ca II}$ K line. These galaxies host a
mixture of old and young stellar populations together with on-going
star formation. Our ground-based imaging does not allow us to securely
determine the morphology of these galaxies, but the spectral energy
distribution (SED) and the spectral features suggest that they may be
Sa-Sb type galaxies. In terms of spectral classification, these
galaxies have a k + a spectrum (Franx 1993) with [OII]
superposed. Based on the equivalent widths of [OII] and $H_{\delta}$
only, these galaxies may be classified as e(c) (Poggianti et
al. 1999). However, their colours are redder than e(c) galaxies and as
red as k type galaxies. Therefore we have decided to classify them as
a + k + [OII] (Table 4) to denote their star-forming nature as well as
their red colour and the post-starburst component.

Similar to the other star-forming galaxies in \CL, these red
star-forming galaxies are located in the outskirts of the cluster (see
Fig. \ref{el_membs}). They are also spatially isolated. The reddest
emission line galaxy in the $R-K_\mathrm{s}$ vs $K_\mathrm{s}$ diagram
of Fig. \ref{rcs} is galaxy ID=270, which has a disturbed, elongated
morphology. In contrast to the other three red star-forming galaxies,
its prominent [OII] ($\lambda3727$) line is likely to be due to
star-formation activity induced by a possible interaction.

The observation of red star-forming galaxies in the field environment
has already been reported in previous works (e.g., Graham \& Dey 1996;
Afonso et al. 2001; Cimatti et al.  2002). Galaxies with red SEDs and
emission lines have been observed in the Canada-France Redshift Survey
(Hammer et al. 1997), and 8\% of these galaxies at $z < 0.7$
can be places of a vigorous star-formation activity affected by strong
dust absorption. In the cluster environment, on the other hand, Duc et
al. (2002) found that about 9\% of galaxies in the CRS of Abell 1689, have
a star-formation activity which is mainly detected by their observed
flux at 15 $\mu$m, with all but one of these galaxies showing no
measurable [OII] ($\lambda$3727). This indicates that most of the
star-formation activity in this $z=0.18$ cluster is hidden. These
obscured star-forming galaxies have projected distances within 550 kpc
from the center of the cluster. In addition to this, all but one of
the star-forming galaxies in the CRS of \object{Abell 1689} are classified as k type
(Poggianti et al. 1999). At lower redshift, the CMD of the Virgo
cluster presented by Gavazzi et al. (2002) shows that some luminous
spiral galaxies have colours as red as cluster ellipticals, however no
explicit indication of their star-forming nature is given.

Although no red k + a with [OII] galaxies have been reported in the
distant cluster \object{MS1054} (e.g., van Dokkum et al. 2000) or the
super cluster \object{Cl 1604+4304} (e.g., Postman, Lubin \& Oke 1998;
Gal \& Lubin 2004), there is evidence that some galaxies with these
peculiar spectro-photometric characteristics have indeed been detected
in RDCS \object{J0848+4453} (van Dokkum \& Stanford 2003), a cluster
at z=1.27 in the Lynx field (Stanford et al. 1997; Rosati et
al. 1999). Two of the three galaxies presented in van Dokkum \&
Stanford (2003) have prominent [OII] ($\lambda$3727) in emission,
accompanied by some evidence for enhanced H$_{\delta}$ ($\lambda$4102)
in absorption. These galaxies have $I-H > 3$ colours and, in contrast
to \CL, are observed in the central 1.1 Mpc $\times$ 1.1 Mpc region of
\object{RDCS J0848+4453}. The X-ray morphology of RDCS J0848+4453
(Stanford et al. 2001) is as irregular as that of \CL. However, the
X-ray morphologies of \object{MS1054} (Gioia et al. 2004) and
\object{Cl 1604+4304} (Lubin, Mulchaey \& Postman 2004) are also
irregular, so a connection between the existence and location of this
type of galaxy in a cluster and the cluster dynamical state is not
clear.

The observation of galaxies with a mixture of stellar populations in a
forming cluster at high redshift represents a key point in the study
of galaxy evolution, since these galaxies may be transition objects
between field spirals and cluster elliptical galaxies.

\begin{center}
\begin{figure}[t]
\centerline{\psfig{file=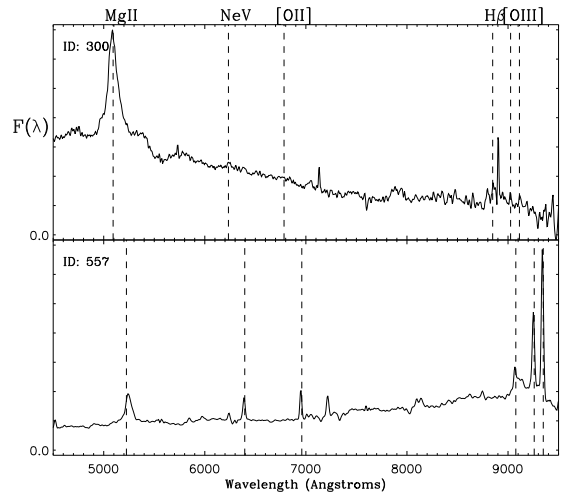,width=8cm}}
\caption{Spectra of the two AGNs that are cluster members. The flux is
in arbitrary units.  From top to bottom the AGNs are at z=0.8201
(ID=300) and z=0.8672 (ID=557). The dashed vertical lines
indicate some important emission features. From left to right these
are MgII ($\lambda$2798), NeV ($\lambda$3425), [OII] ($\lambda$3727),
H$_{\beta}$ ($\lambda$4861), [OIII] ($\lambda$4959) and [OIII]
($\lambda$5007).}
\label{agn0152}
\end{figure}
\end{center}

\begin{center}
\begin{figure}[t]
\centerline{\psfig{file=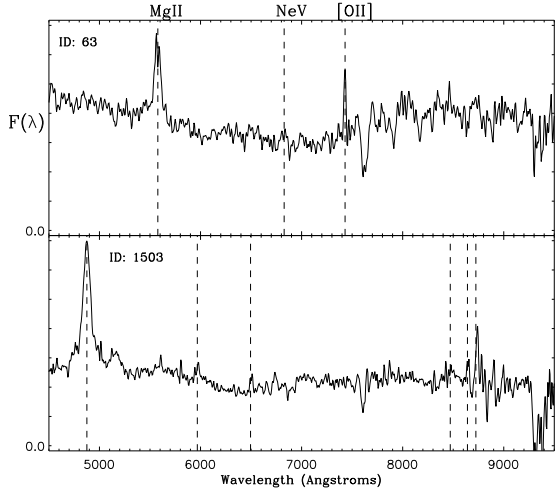,width=8cm}}
\caption{Spectra of the two AGNs that are not cluster members. The flux
is in arbitrary units. From top to bottom the AGN are at z=0.993
(ID=63) and z=0.7439 (ID=1503) respectively. The dashed vertical lines
indicate some important emission features (same as in
Fig. \ref{agn0152}).}
\label{agn_nomembs}
\end{figure}
\end{center}

\begin{table*}[t]
\label{tab_agns}
\begin{center}
\begin{tabular}{c|ccccccc}
\hline
{\small ID} & $\alpha$ (h:m:s) & $\delta$ ($^o$:${'}$:${''}$) & {\small z}& {\small $f_X^{(*)}$} & {\small $L_X^{(**)}$} & {\small $f_X^{(*)}$} & {\small $L_X^{(**)}$} \\
{\small } & {\small (J2000)} & {\small (J2000)} & & {\small [0.5-2] keV} & {\small [0.5-2] keV} & {\small [2-10] keV} & {\small [2-10] keV} \\
\hline
\hline
{\small 300} & {\small 01:52:43.738} & {\small -13:59:01.39} & 0.8201 & {\small 3.5097} & {\small 1.0540} & {\small 3.5989} & {\small 1.2777} \\
{\small 557} & {\small 01:52:39.780} & {\small -13:57:40.10} & 0.8672 & {\small 1.6961} & {\small 0.1529} & {\small 7.6439} & {\small 1.9623} \\
\hline 
\hline
{\small 63} & {\small 01:52:40.915} & {\small -14:00:09.72} & 0.9934 & {\small 0.5546} & {\small 0.2362} & {\small 1.1461} & {\small 0.4696} \\
{\small 1503} & {\small 01:52:34.676} & {\small -13:59:30.54} & 0.7439 & {\small 0.3881} & {\small 0.0492} & {\small 0.9492} & {\small 0.2043} \\
\hline
\end{tabular}
\caption{X-ray fluxes and luminosities of the spectroscopically
observed AGN in the FORS1 field of view of \CL. The first two objects
are confirmed cluster members, the other two are field AGNs. All
fluxes were computed assuming no intrinsic absorption and a galactic
absorption of ${n_H}_{gal}$=1.54 $\times$ 10$^{22}$ cm$^{-2}$
frozen. The spectral slopes, $\gamma$=1.8-2, are typical for type-1
AGN. Errors (1-$\sigma$) on $L_X$ are of 20\%, 25\%, 22\% and 70\% for
IDs 300, 557, 63 and 1503 respectively.(*): Fluxes are in units of
10$^{-14}$ erg s$^{-1}$ cm$^{-2}$. (**): Luminosities are in units of
10$^{44}$ erg s$^{-1}$.  }
\end{center}
\end{table*}

\subsection{X-ray point sources}

The high angular resolution of Chandra has allowed us to identify for
the first time a number of X-ray point sources in the FORS1 field of
\CL. Previous observations of \CL\ with ROSAT and BeppoSAX (Della Ceca
et al., 2000; Romer et al., 2000; Ebeling et al., 2000) were unable to
disentangle some of these X-ray point sources from the diffuse
emission from the ICM, due to the limited spatial resolution of those
satellites. Four of the point sources were targeted during our survey,
and all of them are AGNs.

Two of the four AGNs (ID=300 and ID=557), with redshifts of z=0.8201
and z=0.8672 respectively, are cluster members. One AGN, ID=300, is
located $\sim$ 1\farcm16 ($\sim$ 530 kpc) to the southeast of the
brightest cluster galaxy in the southern clump (ID=387). The other
AGN, ID=557, is located $\sim$46\arcsec\ ($\sim$ 350 kpc) north of
ID=387 on the line dividing the two main interacting clumps. The
location of these AGNs is indicated by the arrows in
Fig. \ref{gal_x_dist}. The number of net counts in the [0.5-10] keV
band are 314.4 and 233 for ID=300 and ID=557 respectively. Their X-ray
fluxes and luminosities obtained from the spectroscopic analysis of
the Chandra data are given in Table \ref{tab_agns}. The spectra are
shown in Figure \ref{agn0152} and both have a prominent broad MgII
($\lambda$2798) emission line characteristic of AGN.

It is interesting to note the presence of several AGN in the cluster
field of view and the fact that the two X-ray point sources closest to
the central double core region are also cluster members. These
observations are consistent with a picture in which AGN are
considered tracers of large scale structures in the universe (e.g.,
Gilli et al. 2003), however they tend to avoid the central regions of
clusters. More AGN identifications around clusters are needed to
bridge the gap between moderate density environments (filaments) and
high-density regions (clusters).

The other two AGNs (ID=63 and ID=1503 in Table \ref{tab_field}
and Fig. \ref{el_membs}) have redshifts of z=0.9934 and z=0.7439,
respectively, and are not cluster members. Their net counts in the
[0.5-10] keV band are 53.197 and 44.019 respectively. Their X-ray
fluxes and luminosities are given in Table \ref{tab_agns}.

The cluster X-ray flux measured by ROSAT PSPC in the [0.5-2] keV band
within a 3 arcmin radius aperture is $f_x=(2.2\pm0.2)\times 10^{-13} \
erg \ s^{-1} \ cm^{-2}$ (Della Ceca et al. 2000). The four X-ray point
sources reported in this section are located within the above
aperture, however ID=1503 was masked out in the PSPC flux estimation
(Della Ceca et al. 2000). From the X-ray fluxes presented in Table
\ref{tab_agns}, we estimate that the ROSAT flux of \CL\ in the [0.5-2]
keV band was overestimated by about 35\% due to contamination from the
other three AGNs.

\section{Discussion and Conclusions}

The extensive spectroscopic survey that we have carried out on \CL\
allows us to firmly characterize its dynamical state. The coverage of
the cluster field was performed in the most uniform possible manner,
allowing us to spectroscopically confirm 102 cluster members.  The
picture derived from this sample confirms previous findings from X-ray
observations. Cluster galaxies are observed to form substructures that
coincide with those in the extended X-ray emission. Furthermore, weak
lensing data based on ACS imaging (Jee et al. 2004) show that the
three mayor clumps of baryons (hot gas and galaxies) are associated to
three separated dark matter halos.

The simple analysis of the cluster galaxy distribution in velocity
space presented in section \S\ref{gal_dist} gives a velocity
dispersion of $\sim$~1600 km s$^{-1}$ to the whole cluster
structure. This value is somewhat higher than that measured for MS1054
(Tran et al. 1999) at z=0.833 and almost the same, within the
uncertainties, to the one measured for RX J1716.6+6708 (Gioia et
al. 1999), another dynamically young galaxy cluster at z=0.81. This
overall velocity dispersion turns out to be larger than that expected
from the $\sigma-T$ relation, indicating the unrelaxed state of the
cluster. The same analysis allowed us to estimate the velocity
dispersion for both the northern and southern clumps, which are in
agreement with the observed $\sigma-T_X$ (Wu, Fang \& Xu 1998) and
$L_x-\sigma$ (Xue \& Wu 2000) relations for clusters. In terms of
total cluster mass, it is not possible to give a reliable estimate
based on the observed velocity dispersion due to the unvirialised
state of the cluster. Numerical simulations (see Roettiger, Loken \&
Burns 1997) show that mergers make the gas evolve differently from the
dark matter, affecting the hydrostatic equilibrium of the gas and,
therefore, the dynamical mass estimates based on the cluster X-ray
emission as well. The evidence presented in this work as well in
others (Maughan et al. 2003; Jee et al. 2004) shows that this is the
case for \CL. In fact, the predicted $M-\sigma$ relation for galaxy
clusters (Bryan \& Norman 1998) is unable to reconcile the weak
lensing mass (Jee et al. 2004) with the overall velocity dispersion
presented in this paper. A velocity dispersion of about 1600 km
s$^{-1}$ would correspond to a virial mass more than a factor of two
greater than the weak lensing measurement. In the case of the
northern and southern subclusters, our mass estimates from
Eq. \ref{virmass} are in close agreement, within the uncertainties,
with those from the weak lensing and X-ray analyses for the same
substructures and within apertures of $\sim$400 kpc in radius.

The spectrophotometric information of a significant fraction of
galaxies down to $\sim R^*+1$ has allowed us to characterize the
galaxy populations in \CL. In terms of spectroscopic classification,
we found the expected spectral type--density relation. The high
density regions of the cluster are dominated by red passive galaxies,
most of them classified as k type. The lower density regions in the
cluster periphery, on the other hand, are dominated by star-forming
[OII] galaxies. All galaxies showing on-going star-forming activity
clearly avoid the high-density regions as traced by the X-ray gas. We
note that star-forming galaxies that are the closest to the central
double core structure of \CL\ are located at $\sim$ 230 kpc away from
the center of the southern clump. This distance is consistent with the
location at which starvation (slow decrease in the star formation rate
due to ram-pressure stripping, thermal evaporation or turbulent and
viscous stripping) starts to become important (see Treu et al. 2003,
and references therein).

An important result is the observation of a subpopulation of galaxies
belonging to the cluster red sequence ($R-K_s \simeq 5$) which are
characterized by the presence of a post-starburst stellar population
together with on-going star-formation. Such galaxies, composed of a
mixture of stellar populations of different ages, point toward a
complex galaxy evolution in such a massive forming cluster. These red
star-forming k + a galaxies may be a transition stage from late-type
field galaxies to early-type galaxies in high-density
regions. Galaxies with post-starburst and on-going star-formation
spectral features have been observed in the core of the distant
cluster RDCS J0848+4453 ($z=1.27$; van Dokkum \& Stanford 2003),
however the strength of the [OII] and Balmer-line features in the
spectrum of galaxies ID=248 and ID=551 indicate that more extensive
star-forming activity may be occurring in a disk structure. These red
star-forming galaxies might represent a cluster elliptical caught in
the act of formation. The link between this population and the
dynamical stage of the cluster is something that requires further
investigation.

We conclude that the velocity dispersion estimates, indicate that \CL\
is a massive system already in place when the universe was 6.46 Gyr
old (48\% of its present age). The combination of all the data
available on \CL\ leads us to conclude that in this system the dark
matter and the baryons trace each other very well, indicating an
advanced stage of thermalisation in the cluster potential well. The
filamentary structure observed in the \CL\ galaxy distribution agrees
well with the X-ray distribution of its ICM. There is strong evidence
for significant substructure in this system, with two major
sub-clusters in a merging phase. \CL\ shows hierarchical structure
formation caught in the act, indicating that large structures are
produced by the merging of subunits in the earlier phases of
evolution, as expected from the classical picture of hierarchical
clustering. \CL\ is an ideal laboratory, where processes affecting
galaxy formation are ongoing. It provides us with an opportunity to
improve our understanding of galaxy evolution in clusters.

\begin{acknowledgements}

This work would not have been possible without the dedicated efforts
of ESO staff, in both Chile and Europe. R.D is grateful to
Drs. H. Flores and N. Cross for interesting and useful discussions. We
are grateful to Dr. Tadayuki Kodama for providing us with a copy of
his models and to Dr. John Blakeslee for useful comments and for
kindly running galaxy evolution models. We are grateful to the
anonymous referee for helpful comments. We wish to acknowledge the
great cultural and spiritual role that the summit of Mauna Kea has
within the indigenous Hawaiian community and express our gratitude for
permission to observe from its summit. We are grateful to the Keck
observatory staff for the support during the observations. The
W. M. Keck observatory is a scientific partnership between the
University of California and the California Institute of Technology,
made possible by the generous gift of the W. M. Keck Foundation.

\end{acknowledgements}


\begin{thebibliography}{}

\bibitem {} Afonso, J. et al. 2001. ApJL, 559, 101.
\bibitem {} Appenzeller , I. \& Rupprecht, G. The ESO Messenger, March 1992. p.18.
\bibitem {} Beers, T. C. et al. 1990. AJ, 100, 32.
\bibitem {} Ben\'{\i}tez, N. 2000. ApJ, 536, 571.
\bibitem {} Ben\'{\i}tez, N. et al. 2004. ApJS, 150, 1. 
\bibitem {} Bertin, E. \& Arnouts, S. 1996. A\&AS, 117, 393.
\bibitem {} Bryan, G. L. \& Norman, M. L. 1998. ApJ, 495, 80.
\bibitem {} Bruzual A., G. \& Charlot, S. 1993. ApJ, 405, 538.
\bibitem {} Cimatti, A. et al. 2002. A\&A, 381, L68.
\bibitem {} Coleman, G. D., Wu, C. \& Weedman, D. W. 1980. ApJS, 43, 393.
\bibitem {} Cross, N. J. G., et al. 2004. AJ, in press.
\bibitem {} Della Ceca, R. et al. 2000. A\&A, 353, 498.
\bibitem {} Donahue, M. et al. 1998. ApJ, 502, 550.
\bibitem {} Dressler, A. \& Gunn, J. E. 1983. ApJ, 270, 7.
\bibitem {} Dressler, A. \& Schectman, S. 1988. AJ, 95, 985.
\bibitem {} Dressler, A. et al. 1999. ApJSS, 122, 51.
\bibitem {} Duc P.-A. et al. 2002. A\&A, 382, 60. 
\bibitem {} Ebeling, H. et al. 2000. ApJ, 534, 133.
\bibitem {} Ettori, S. et al. 2004. A\&A, 417, 13.
\bibitem {} Franx, M. 1993. ApJL, 407, 5.
\bibitem {} Gal, R. R. \& Lubin, L. M. 2004. ApJL, 607, 1.
\bibitem {} Gavazzi, G. et al. 2002. ApJ, 576, 135.
\bibitem {} Gilli, R. et al. 2003. ApJ, 592, 721.
\bibitem {} Gioia, I. M. et al. 1999. AJ, 117, 2608.
\bibitem {} Gioia, I. M. et al. 2004. A\&A, 419, 517.
\bibitem {} Graham, J. R. \& Dey, A. 1996. ApJ, 471, 720.
\bibitem {} Hammer, F. et al. 1997. ApJ, 481, 49.
\bibitem {} Homeier, N. L. et al. 2004. ApJ, in press.
\bibitem {} Huo, Z. et al. 2004. AJ, 127, 1263.
\bibitem {} Jee, M. J. et al. 2004. ApJ, in press.
\bibitem {} Joy, M. et al. 2001. ApJL, 551, 1.
\bibitem {} Kinney, A. L. et al. 1996. ApJ, 467, 38.
\bibitem {} Kodama, T. \& Arimoto N. 1997. A\&A, 320, 41.
\bibitem {} Kurtz, M. J. et al. 1992. ADASS, 1, 432.
\bibitem {} Longair, M. S. 1998. Galaxy Formation. A\&A library. Springer-Verlag.
\bibitem {} Lubin, L. M., Mulchaey, J. S. \& Postman, M. 2004. ApJL, 601,~9.
\bibitem {} Maughan, B. J. et al. 2003. ApJ, 587, 589.
\bibitem {} Mercurio, A. et al. 2003a. A\&A, 397, 431.
\bibitem {} Moorwood, A., Cuby, J.-G. \& Lidman, C. 1998. ESO Messenger, 91, 9
\bibitem {} Oke, J. B. et al. 1995. PASP, 107, 375.
\bibitem {} Poggianti, B. M. et al. 1999. ApJ, 518, 576.
\bibitem {} Postman, M., Lubin, L. M. \& Oke, J. B. 1998. AJ, 116, 560.
\bibitem {} Roettiger, K., Loken, C. \& Burns, J. O. 1997. ApJS, 109, 307.
\bibitem {} Romer, A. K. et al. 2000. ApJSS, 126, 209.
\bibitem {} Rosati, P. et al. 1998. ApJL, 492, 21.
\bibitem {} Rosati, P. et al. 1999. AJ, 118, 76.
\bibitem {} Rose, J. A. 1984. AJ, 89, 1238.
\bibitem {} Rose, J. A. 1985. AJ, 90, 1927.
\bibitem {} Stanford, S. A. et al. 1997. AJ, 114, 2232.
\bibitem {} Stanford, S. A. et al. 2001. ApJ, 552, 504.
\bibitem {} Tonry, J. \& Davis, M. 1979. AJ, 84, 1511.
\bibitem {} Tran, K. H. et al. 1999. ApJ, 522, 39.
\bibitem {} Treu, T. et al. 2003. ApJ, 591, 53.
\bibitem {} van Dokkum, P. G. et al. 1999. ApJL, 520, 95.
\bibitem {} van Dokkum, P. G. et al. 2000. ApJ, 541, 95.
\bibitem {} van Dokkum, P. G. \& Stanford, S. A. 2003. ApJ, 585, 78.
\bibitem {} Wu, X., Fang, L. \& Xu, W. 1998. A\&A, 338, 813.
\bibitem {} Xue, Y. \& Wu, X. 2000. ApJ, 538, 65.

\end{thebibliography}
\end{document}